\documentclass[12pt]{article}

\usepackage{amsmath}
\usepackage{amscd}
\usepackage{amsfonts}
\usepackage{epsfig}
\usepackage{theorem}
\usepackage{authblk}
\usepackage[titletoc]{appendix}
\usepackage[margin=1.0in]{geometry}

\DeclareMathOperator{\rect}{rect}

\title{Characterizing the Narrowband Ambiguity Function of Multi-Tone Sinusoidal Frequency Modulated Waveforms}

\author{David A.~Hague}

\affil{Sensors and Sonar Systems Department \\ Naval Undersea Warfare Center\\\texttt{david.a.hague.civ@us.navy.mil}}

\begin{document}
\maketitle
\hspace{-0.6cm}
  
\vspace{24pt}

\textit{\hspace{-0.6cm}
\textbf{Abstract:}
This paper characterizes the mainlobe and sidelobe structure of the Multi-Tone Sinusoidal Frequency Modulated (MTSFM) transmit waveform's narrowband Ambiguity Function (AF) for active sonar applications. The MTSFM waveform's modulation function is represented as a Fourier series. The Fourier coefficients form a discrete set of parameters that are modified to synthesize waveforms with novel characteristics. The contour of the AF’s mainlobe is well approximated as a coupled ellipse known as the Ellipse Of Ambiguity (EOA). The EOA parameters determine the mainlobe width in range and Doppler as well as the degree of coupling between them. This coupling factor, known as the Range-Doppler Coupling Factor (RDCF), determines whether a waveform is Doppler sensitive or Doppler tolerant. This paper derives exact closed form expressions for the EOA parameters of the MTSFM's AF. The MTSFM's design coefficients allow for fine control of its AF mainlobe width in range and Doppler as well as its RDCF. This fine control facilitates designing waveforms that can smoothly trade-off between possessing Doppler sensitive and Doppler tolerant characteristics. Additionally, this paper introduces a method to control the sidelobe structure of the MTSFM's Auto Correlation Function (ACF) while maintaining the waveform's AF mainlobe shape. This is achieved using a numerical optimization technique that minimizes the ratio of $\ell_2$-norms of the ACF mainlobe and sidelobe regions subject to constraints on the EOA parameters. Simulations demonstrate the effectiveness of this optimization technique.}
\vspace{24pt}

\textit{\hspace{-0.6cm}
}

\newpage

\section{Introduction}
\label{sec:Intro}
Waveform diversity, the ability to optimize waveforms to maximize system performance, possibly in a dynamically adaptive manner, has been an active topic of research in the radar community for over two decades \cite{Blunt_Waveform_Diversity}.  This field of research has been enabled by the development of several parameterized modulation techniques such as Phase-Coding (PC) and Frequency Shift-Keying (FSK) \cite{Levanon} which facilitate the synthesis of waveforms with novel characteristics.   Waveform diversity has increasingly become a topic of interest to the active sonar community with diverse sets of waveforms being employed for Multi-Beam Echo Sounding (MBES) \cite{MBES_Orthogonal_Waveforms} and Multiple-Input Multiple Output (MIMO) sonar applications \cite{Hansen_MIMO_SAS}.  These efforts highlight the need for continued development of parameterized waveform models to further enable waveform diversity for active sonar systems.  

Recently, the author developed the Multi-Tone Sinusoidal Frequency Modulated (MTSFM) waveform as a novel parameterized FM waveform model \cite{Hague_IEEE_AES}.  The MTSFM waveform's phase and frequency modulation functions are composed of a finite Fourier series.  The Fourier coefficients representing the waveform's instantaneous phase and frequency functions are utilized as a discrete set of adjustable parameters.  To date, most efforts have focused on the design of Doppler sensitive MTSFM waveforms that possess a ``Thumbtack-Like'' Ambiguity Function (AF).  However, the parameterization of the MTSFM model suggests that a variety Doppler tolerant designs should also be feasible.  

This paper describes a method to jointly control the MTSFM waveform's AF mainlobe shape and the sidelobe structure of its Auto Correlation Function (ACF).  The AF mainlobe is characterized using the Ellipse of Ambiguity (EOA) model.  The MTSFM's design coefficients allow for fine control of its AF mainlobe width.  This facilitates synthesizing waveforms that can smoothly trade-off between possessing Doppler sensitive and Doppler tolerant characteristics.  Additionally, this paper demonstrates a numerical optimization technique that minimizes an $\ell_2$ norm on the mainlobe and sidelobe regions of the waveform's ACF subject to constraints on the MTSFM's EOA parameters.  This allows for controlling the sidelobe structure of the MTSFM's ACF while maintaining the waveform's AF mainlobe shape.  The rest of this paper is organized as follows: Section 2 introduces the MTSFM waveform model and the metrics used to describe the AF mainlobe and ACF sidelobe structure; Section 3 demonstrates how the MTSFM's parameters can be utilized to shape its AF mainlobe and reduce ACF sidelobes via several illustrative design examples; finally, Section 4 concludes the paper.

\section{MTSFM Waveform and the Ambiguity Function}
\label{sec:Model}
In general, a basebanded FM waveform is expressed in continuous time as
\begin{equation}
s\left(t\right)=\frac{\rect\left(t/T\right)}{\sqrt{T}} e^{j\varphi\left(t\right)}
\label{eq:complexExpo}
\end{equation}
where $T$ is the waveform's duration and $\varphi\left(t\right)$ is its phase modulation function. The $1/\sqrt{T}$ term normalizes the signal energy to unity.  The waveform's frequency modulation function $m\left(t\right)$ maps its instantaneous frequency as a function of time and is expressed as $m\left(t\right) = \left(1/2\pi\right)\partial\varphi(t)/\partial t$.
The MTSFM waveform's frequency and phase modulation functions are expressed as a finite Fourier series
\begin{align}
m\left(t\right) &= \frac{a_0}{2}+\sum_{\ell=1}^L a_{\ell}\cos\left(\dfrac{2\pi \ell t}{T}\right) + b_{\ell}\sin\left(\dfrac{2\pi \ell t}{T}\right), \label{eq:MTSFM_Mod} \\
\varphi\left(t\right) &= \pi a_0 t + \sum_{\ell=1}^L\alpha_{\ell}\sin\left(\dfrac{2\pi \ell t}{T}\right) - \beta_{\ell}\cos\left(\dfrac{2\pi \ell t}{T}\right)
\label{eq:MTSFM_Phase}
\end{align}
where $a_{\ell}$ and $b_{\ell}$ are the Fourier coefficients and $\alpha_{\ell} = a_{\ell}T/\ell$ and $\beta_{\ell} = b_{\ell}T/\ell$ are the waveform's modulation indices.  The Fourier coefficients of the MTSFM waveform's modulation function, and correspondingly, the modulation indices of the waveform's instantaneous phase, form a discrete set of adjustable parameters that can be modified to synthesize waveforms with a variety of unique properties \cite{Hague_IEEE_AES, Hague_SSP_2018}.  The narrowband Ambiguity Function (AF) measures the response of the waveform's MF to its Doppler shifted versions and is defined as \cite{Ricker}
\begin{equation}
\chi\left(\tau, \nu\right) = \int_{-\infty}^{\infty}s\left(t\right)s^*\left(t+\tau\right)e^{j2\pi \nu t} dt
\label{eq:AF}
\end{equation}
where $\nu$ is the Doppler frequency shift.  Lastly, the ACF is the zero Doppler cut of the AF $R\left(\tau\right) = \chi\left(\tau, 0\right)$.  There are a series of design metrics that characterize the mainlobe and sidelobe structure of a waveform's AF and ACF shape.  These metrics are described below.

\subsection{Mainlobe Structure}
\label{subsec:mainlobe}
The AF mainlobe structure determines a waveform's ability to estimate the range and Doppler of a target and to resolve multiple targets in range and Doppler.  The AF mainlobe can be approximated by a second order Taylor series expansion \cite{Ricker}.  The EOA is the contour of the mainlobe approximation at some height $1-\xi$ which is always a coupled ellipse and is expressed as 
\begin{equation}
1 - |\chi\left(\tau, \nu\right)|^2 = \xi = \beta_{rms}^2\tau^2 + 2\rho\tau\nu + \tau_{rms}^2\nu^2
\end{equation}
where $\beta_{rms}^2$ is the waveform's RMS bandwidth and determines time-delay (range) sensitivity, $\tau_{rms}^2$ is the RMS pulse length which determines Doppler sensitivity, and $\rho$ is the RDCF for the AF mainlobe.  The RMS bandwidth is expressed as \cite{Ricker}
\begin{equation}
\beta_{rms}^2 = \left(2\pi\right)^2\int_{\infty}^{\infty}\left(f - f_0\right)^2 |S\left(f\right)|^2 df = \dfrac{1}{T}\int_{-T/2}^{T/2} \left[\dot{\varphi}\left( t \right)\right]^2 dt - \left| \dfrac{1}{T}\int_{-T/2}^{T/2} j \dot{\varphi}\left(t\right) dt \right| ^2
\label{eq:Brms}
\end{equation}
where $f_0$ is the waveform's spectral centroid $\langle f \rangle$, $S\left(f\right)$ is the waveform's Fourier transform, and $\dot{\varphi}\left(t\right)$ is the first time derivative of the waveform's instantaneous phase.  The RMS pulse length term is expressed as \cite{Ricker}
\begin{equation}
\tau_{rms}^2 = 4\pi^2\int_{\Omega_t} \left(t-t_0\right)^2 |s\left(t\right)|^2 dt 
\label{eq:Trms}
\end{equation}
where $t_0$ is the first time moment $\langle t \rangle$ of the waveform's complex envelope $|s\left(t\right)|^2$ and is zero for waveforms such as \eqref{eq:complexExpo} that are even-symmetric in time.  The RDCF $\rho$ is expressed as \cite{Ricker}
\begin{align}
\rho &= -2\pi \Im \Biggl\{\int_{\Omega_t} ts\left(t\right)\dot{s}^*\left(t\right) dt \Biggr\}  =2\pi  \int_{-T/2}^{T/2}  t  \dot{\varphi}\left(t\right) dt
\label{eq:gamma}
\end{align}
where $\Im \{\}$ denotes the imaginary component of the integral. The RMS bandwidth and RDCF EOA parameters are solely dependent upon the frequency modulation function $\dot{\varphi}\left(t\right)$.  If the waveform's modulation function is known, $\beta_{rms}^2$ and $\rho$ can be calculated in exact closed form which along with $\tau_{rms}^2$ provides full characterization of the waveform's AF mainlobe structure \cite{Ricker}.  Furthermore, if the waveform's modulation function is composed of a discrete set of parameters, those parameters can be optimized to design waveforms with specific EOA parameters thus uniquely shaping the waveform's AF mainlobe structure \cite{Hague_SSP_2018}.

\subsection{Sidelobe Structure}
\label{subsec:sidelobe}
This paper focuses specifically on the sidelobe structure of the ACF.  Two of the most common metrics are the Peak-to-Sidelobe Level Ratio (PSLR) and the Integrated Sidelobe Level (ISL) \cite{Blunt_Waveform_Diversity}.  The PSLR is expressed as
\begin{equation}
\text{PSLR} =  \dfrac{\underset{\Delta\tau \leq \vert\tau\vert \leq T}{\text{max}}\bigl\{\left|R\left(\tau\right)\right|^2\bigr\}}{\underset{0 \leq \vert\tau\vert \leq \Delta \tau}{\text{max}}\bigl\{\left|R\left(\tau\right)\right|^2\bigr\}} = \underset{\Delta \tau \leq \vert\tau\vert \leq T}{\text{max}}\bigl\{\left|R\left(\tau\right)\right|^2\bigr\}
\label{eq:PSLR}
\end{equation} 
where $\Delta \tau$ is the null of the ACF mainlobe thus establishing the ACF's null-to-null mainlobe width as $2\Delta \tau$.  Note that the rightmost expression in \eqref{eq:PSLR} results from the assumption that the waveform is unit energy and thus the maximum value of $|R\left(\tau\right)|^2$ is unity which occurs at $\tau = 0$.  The ISL is the ratio of the $\ell_2$-norms of the mainlobe and sidelobe regions of the ACF expressed as
\begin{equation}
\text{ISL}~= \dfrac{\int_{\Delta \tau}^{T}\left|R\left(\tau\right)\right|^2 d\tau}{\int_{0}^{\Delta \tau}\left|R\left(\tau\right)\right|^2 d\tau}.
\label{eq:ISL}
\end{equation}
Note that the integration is performed only over positive time-delays since the ACF is even-symmetric in $\tau$.  A lower ISL  corresponds to an ACF with lower overall sidelobe levels but does not necessarily translate to a lower PSLR.

\section{Shaping the MTSFM's AF Mainlobe and Sidelobe Structure}
\label{sec:bafMTSFM}
The MTSFM's EOA parameters are expressed in terms of the Fourier coefficients in its frequency modulation function as \cite{Hague_IEEE_AES, Hague_SSP_2018}
\begin{align}
\beta_{rms}^2 &= 2\pi^2\sum_{\ell}^L  a_{\ell}^2 + b_{\ell}^2, \label{eq:Brms1}\\
\tau_{rms}^2 &= \dfrac{\pi^2T^2}{3}, \label{eq:Trms1}\\
\rho &= -2\pi^2 T\sum_{\ell=1}^L b_{\ell} \dfrac{\cos\left(\pi \ell\right)}{\pi \ell} . \label{eq:rho1}
\end{align}
It's important to note several observations regarding the results shown in \eqref{eq:Brms1}-\eqref{eq:rho1}.  First, as with all rectangularly tapered waveforms, the Doppler sensitivity is proportional to the square of the waveform duration $T$ times a constant ($\pi^2/3$).  Second, both \eqref{eq:Brms1} and \eqref{eq:rho1} are controlled by the MTSFM's coefficients.  Lastly, the RDCF \eqref{eq:rho1} is a function only of the odd harmonic coefficients in \eqref{eq:MTSFM_Mod}.  This is consistent with the results in \cite{Ricker} describing how waveforms with even symmetric modulation functions possess an AF mainlobe with RDCF of zero.  

\subsection{The MTSFM's AF Mainlobe Structure - A Simple Design Example}
\label{subsec:synthesisMainlobe}
\begin{figure}[ht]
\centering
\includegraphics[width=0.75\textwidth]{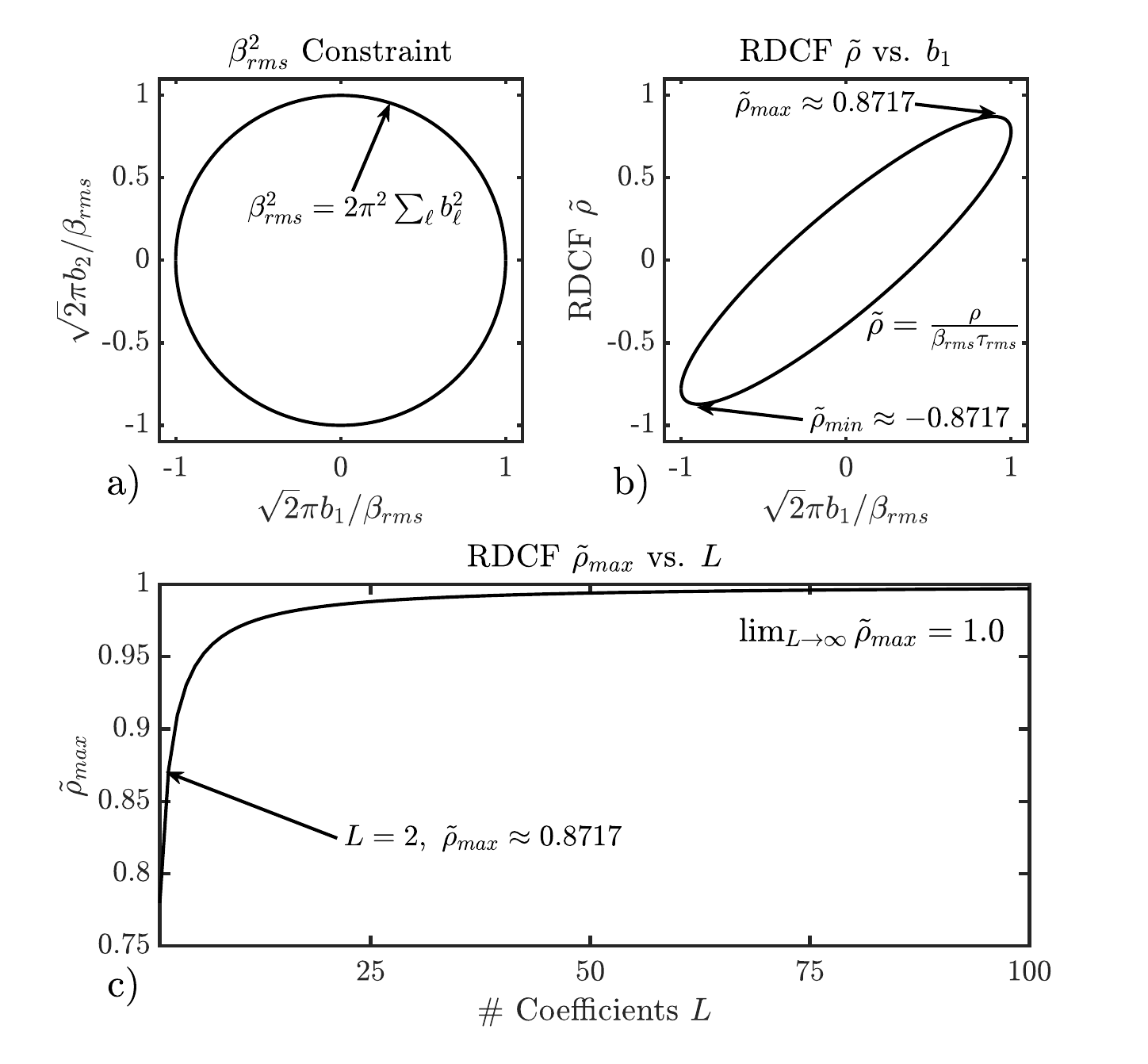}
\caption{MTSFM coefficients that satisfy the RMS bandwidth equality constraint (a), the corresponding RDCF values $\tilde{\rho}$ of these MTSFM coefficients (b), and the optimal $\tilde{\rho}_{\text{max}}$ for a MTSFM with $L$ harmonics in its modulation function (c).}
\label{fig:MTSFM1}
\end{figure}

Consider the following design examples that synthesize MTSFM waveforms with two sine Fourier coefficients $b_1$ and $b_2$ with an RMS bandwidth equality constraint.  Using \eqref{eq:Brms1}, the second coefficient $b_2$ can be expressed in terms of the first coefficient $b_1$ as 
\begin{equation}
b_2 = \pm \sqrt{\left(\dfrac{\beta_{rms}^2}{2\pi^2} - b_1^2\right)}.
\label{eq:MTSFM_b1}
\end{equation}
The RDCF $\rho$ is expressed as
\begin{align}
\rho &= \pi T \left[2b_1 - b_2 \right] = \pi T \left[2b_1 \mp \sqrt{\left(\dfrac{\beta_{rms}^2}{2\pi^2} - b_1^2\right)} \right].
\label{eq:MTSFM_gamma_b1}
\end{align}
Figure \ref{fig:MTSFM1} displays $b_2$ and $\rho$ as a function of $b_1$.  Note that $\rho$ is normalized to $\tilde{\rho} = \rho/\left(\beta_{rms}\tau_{rms}\right)$ so as to mimic a correlation coefficient.  This normalized RDCF $\tilde{\rho}$ takes on the values $-1\leq \tilde{\rho} \leq 1$ where $\pm1$ corresponds to an AF mainlobe that is perfectly negatively and positively coupled respectively.  The values of $b_1$ and $b_2$ that satisfy the RMS bandwidth equality constraint form a circle and the corresponding RDCF values $\tilde{\rho}$ form a coupled ellipse.  With only two MTSFM coefficients, $\tilde{\rho}$ takes on all the values between $\approx \pm 0.8717$.  As shown in the Appendix, the $L$ MTSFM coefficients that maximize $\tilde{\rho}$ are
\begin{equation}
b_{\ell} = \dfrac{-\sqrt{2}\beta_{rms}}{2\pi\sqrt{\sum_{\ell'}\left[\dfrac{\cos\left(\pi \ell' \right)}{\pi \ell'}\right]^2}}\dfrac{\cos\left(\pi \ell \right)}{\pi \ell}.
\label{eq:maxRho}
\end{equation}
The resulting $\tilde{\rho}_{\text{max}}$ using these optimal MTSFM coefficients for increasing $L$ are shown in panel (c) of Figure \ref{fig:MTSFM1}.  The MTSFM is capable of possessing essentially any realizable RDCF $-1 < \tilde{\rho} < 1$.  Waveforms with variable RDCF values could have practical applications in active sonar systems.  Doppler sensitive waveforms with long duration require a large number of Doppler shifted matched filters to process Doppler shifted target echoes.  This can substantially increase the complexity of the sonar receiver \cite{Doisy}.  Utilizing a waveform with variable Doppler sensitivity could smoothly trade off between Doppler sensitivity and receiver complexity.

\subsection{Controlling the MTSFM's ACF Sidelobe Structure}
\label{sec:synthesisMainlobe}
The ACF sidelobes of the MTSFM can be reduced while maintaining its AF mainlobe structure.  This is done via optimizing the ISL metric subject to constraints on the RMS bandwidth and RDCF of an initial MTSFM waveform.  Formally, the optimization problem is stated as 
\begin{align}
\underset{b_{\ell}}{\text{min}}
\text{~ISL}\left(\{b_{\ell}\}\right) \text{s.t.} &\left(1-\delta\right)\beta^{(0)}_{rms}\leq \beta_{rms} \leq \left(1+\delta\right)\beta^{(0)}_{rms}, \label{eq:Problem1} \\ &\left(1-\epsilon\right)\rho^{(0)}\leq \rho  \leq \left(1+\epsilon\right)\rho^{(0)} \nonumber
\end{align}
where $\beta^{(0)}_{rms}$ and $\rho^{(0)}$ are the initial seed waveform's RMS bandwidth and RDCF respectively and $\delta = 0.1$ and $\epsilon = 0.05$ are unitless bound parameters.  The following simulation utilizes four waveforms whose initial Fourier coefficients are listed in Table 1.  The waveform coefficients and pulse-lengths were scaled such that the resulting waveforms' RMS bandwidths were equal to that of an LFM with a time-bandwidth product of $T\Delta f = 200$ where $\Delta f$ is the waveform's swept bandwidth.  Note that since waveform III's RDCF was $0.0$, the RDCF constraint was not used during optimization.    These seed MTSFM waveforms were zero-padded with an additional 126 Fourier coefficients to further increase the degrees of freedom in the optimization problem.  This has been shown to produce waveforms with lower ACF sidelobes \cite{Hague_IEEE_AES}.  The optimization problem in \eqref{eq:Problem1} was solved using MATLAB's optimization toolbox and the resulting waveforms' ISL and PSLR values for these simulations are shown in Table 2.  Also shown in Table 2 are the ratios of the optimized waveforms' $\beta_{rms}^2$ and $\rho$ to that of the initial waveforms denoted as $\beta_{rms}/\beta_{rms}^{(0)}$ and $\rho/\rho^{(0)}$ respectively.  All four optimized waveforms exhibit clearly lower ISL and PSLR values.  Additionally, the AF mainlobe EOA parameters of all four waveforms stay relatively close to the values of the initial seed waveforms.  

\section{Conclusion}
\label{sec:Conclusion}
The closed form expressions for the MTSFM waveform's EOA parameters allows for fine control of its resulting AF mainlobe shape.  The Fourier coefficients of the MTSFM specifically allow for finely adjusting the RDCF.  This in turn facilitates designing MTSFM waveforms with adjustable Doppler tolerance introducing a design tradeoff between receiver complexity and Doppler resolution.  Additionally, the sidelobe levels of the resulting waveform's ACF can be reduced by minimizing the ISL.  The RMS bandwidth and RDCF constraints ensure that the resulting optimized waveform will largely retain its AF mainlobe shape.  Future efforts will focus on deriving the MTSFM's broadband AF EOA parameters and extend the optimization routines to shape the broadband AF sidelobes over regions in the range-Doppler plane.  

\begin{table}[htb]

\begin{center}
{
\begin{tabular}{|l||l|l|l|l|l|}\hline
Waveform   & $\sqrt{2}\pi b_1/ \beta_{rms}$      & $\sqrt{2}\pi b_2/ \beta_{rms}$	& $\tilde{\rho}$    \\\hline\hline
         I       &   0.8944									  & -0.4473     						&  0.8717				   \\\hline
         II      &   0.1292									  & -0.9916    				   		     &  0.4873				   \\\hline
         III     &  -0.4472								       & -0.8944     						     &  0.0000				   \\\hline
         IV     &  -0.8944									  &  0.4473     						& -0.8717				   \\\hline
\end{tabular}
\caption[List of MTSFM Waveforms.]{List of MTSFM waveforms with their normalized design coefficients $b_1$ \& $b_2$ and their normalized RDCF $\tilde{\rho}$ taken directly from Figure \ref{fig:MTSFM1}.}
}
\end{center}

\end{table}  

\begin{table}[htb]
\centering
\begin{tabular}{|l||l|l|l|l|}
\hline
Waveform   & Init./Opt. ISL (dB)	& Init./Opt. PSLR (dB)		& $\beta_{rms}/\beta_{rms}^{(0)}$  &  $\rho/\rho^{(0)}$     \\\hline\hline
         I       & -0.01/-15.04  		& -8.89/-28.44			&   1.099     							&      1.011       \\\hline
         II      & 3.91/-10.86   		& -6.19/-21.67   			&   1.099     							&      0.953       \\\hline 
         III     & 0.10/-11.74   		& -11.22/-26.73 			&   1.099     							&      -------        \\\hline
         IV     & 2.54/-14.66    		& -8.0/-30.44			&   1.099     							&      1.018        \\\hline
\end{tabular}
\caption{List of the performance characteristics of the initial and optimized MTSFM waveforms.  The optimized MTSFM waveforms possess substanitally lower PSLR and ISL values than the initial seed waveforms while largely retaining their AF mainlobe shape.}

\end{table}  

\section{Acknowledgements}
This research was supported by the Office of Naval Research (ONR) award \#N0001423WX00345.



\section{Derivation of MTSFM Coefficients that Optimize $\rho$ with an RMS Bandwidth Equality Constraint}
\label{sec:AppendixIV}
The waveform design problem is to find the set of MTSFM coefficients $b_{\ell}$ that maximizes the RDCF $\rho$ subject to an equality constraint on the waveform's RMS bandwidth $\beta_{rms}^2$.   Noting that maximizing $\rho$ is equivalent to minimizing $-\rho$, the optimization problem is formally stated as 
\begin{equation}
\underset{b_{\ell}}{\text{min}}\Biggl\{ 2\pi^2T\sum_{\ell}b_{\ell} \dfrac{\cos\left(\pi \ell\right)}{\pi \ell} \Biggr\}~\text{s.t.}~2\pi^2\sum_{\ell} b_{\ell}^2 = \beta_{rms}^2.
\label{eq:App_II_1}
\end{equation}
Since \eqref{eq:App_II_1} imposes an equality constraint, it can be solved using the method of Lagrange multipliers.  The Lagrangian function is expressed as
\begin{equation}
L\left(b_{\ell}, \lambda\right) =  2\pi^2T\sum_{\ell}b_{\ell} \dfrac{\cos\left(\pi \ell\right)}{\pi \ell} + \lambda\left(2\pi^2\sum_{\ell} b_{\ell}^2 - \beta_{rms}^2\right).
\label{eq:App_II_2}
\end{equation}
Taking the gradient of \eqref{eq:App_II_2} with respect to $b_{\ell}$ and setting equal to $0$ results in the following expressions for each coefficient $b_{\ell}$ in terms of $\lambda$
\begin{equation}
b_{\ell} = \dfrac{-T}{2\lambda}\dfrac{\cos\left(\pi \ell\right)}{\pi \ell}
\label{eq:App_II_3}
\end{equation}
Inserting \eqref{eq:App_II_3} back into the RMS bandwidth constraint results in the expression 
\begin{equation}
\beta_{rms}^2 = 2\pi^2 \sum_{\ell} \left[\dfrac{-T}{2\lambda}\dfrac{\cos\left(\pi \ell\right)}{\pi \ell}\right]^2=\dfrac{2\pi^2T^2}{4\lambda^2}\sum_{\ell} \left[\dfrac{\cos\left(\pi \ell \right)}{\pi \ell}\right]^2.
\label{eq:App_II_4}
\end{equation}
and solving in terms of $\lambda$ yields 
\begin{equation}
\lambda = \dfrac{\pi T}{\sqrt{2} \beta_{rms}}\sqrt{\sum_{\ell} \left[\dfrac{\cos\left(\pi \ell \right)}{\pi \ell}\right]^2}.
\label{eq:App_II_5}
\end{equation}
From here on, the $\sqrt{\sum_{\ell} \left[\cos\left(\pi \ell \right)/\pi \ell\right]^2}$ will be represented in terms of the variable $\ell'$ rather than $\ell$ to avoid confusion with the indices of the MTSFM coefficients $b_{\ell}$.  Inserting the expression for $\lambda$ \eqref{eq:App_II_5} back into the expression for $b_{\ell}$ \eqref{eq:App_II_3} results in the MTSFM design coefficients $b_{\ell}$ that maximize the RDCF $\rho$.  
\begin{equation}
b_{\ell} = \dfrac{-\sqrt{2}\beta_{rms}}{2\pi\sqrt{\sum_{\ell'}\left[\dfrac{\cos\left(\pi \ell' \right)}{\pi \ell'}\right]^2}}\dfrac{\cos\left(\pi \ell \right)}{\pi \ell}.
\label{eq:App_II_6}
\end{equation}
Reassuringly, for the case where $L\rightarrow\infty$, the MTSFM coefficients $b_{\ell}$ become those of an up-sweeping LFM waveform.  When $L\rightarrow\infty$, the square-root term in \eqref{eq:App_II_6} simplifies to $\pi / \sqrt{6}$.  Setting $\beta_{rms} = \pi\Delta f / \sqrt{3}$, the RMS bandwidth of the LFM waveform, results in the expression
\begin{equation}
b_{\ell} = \dfrac{\sqrt{12}\pi \Delta f}{\sqrt{3}2\pi}\dfrac{\cos\left(\pi \ell \right)}{\pi \ell} = \Delta f\dfrac{\cos\left(\pi \ell \right)}{\pi \ell}
\label{eq:App_II_7}
\end{equation}
which are the Fourier coefficients to an up-sweeping LFM waveform's modulation function.

\end{document}